\documentclass[twocolumn,amsmath,amssymb,pra]{revtex4}

\usepackage{graphicx}
\usepackage{dcolumn}
\usepackage{bm}
\usepackage{subfigure}

\begin{document}


\title {
Saddle point anomaly of Landau levels in graphenelike structures
}

\author {A. V. Nikolaev}

\affiliation{
Skobeltsyn Institute of Nuclear Physics, Moscow
State University, Vorob'evy Gory 1/2, 119234, Moscow, Russia
}

\affiliation{Moscow Institute of Physics and Technology, 141700 Dolgoprudny, Russia}

\date{\today}

\begin{abstract}
Studying the tight binding model in an applied rational magnetic field ($H$) we show that in graphene there are very unusual Landau levels
situated in the immediate vicinity of the saddle point ($M$-point) energy $\epsilon_M$.
Landau levels around $\epsilon_M$ are broadened into minibands
(even in relatively weak magnetic fields $\sim40-53$ T) with the maximal width reaching 0.4-0.5 of the energy separation
between two neighboring Landau levels though at all other energies the width of Landau levels is practically zero.
In terms of the semiclassical approach a broad Landau level or magnetic miniband at $\epsilon_M$
is a manifestation of the so called self-intersecting orbit signifying an abrupt transition from
the semiclassical trajectories enclosing the $\Gamma$ point to the trajectories enclosing the $K$ point in the momentum space.
Remarkably, the saddle point virtually does not affect the diamagnetic response of graphene,
which is caused mostly by electron states in the vicinity of the Fermi energy $\epsilon_F$.
Experimentally, the effect of the broading of Landau levels can possibly be observed in twisted graphene
where two saddle point singularities can be brought close to the Fermi energy.
%
%
\end{abstract}



\maketitle

\section{Introduction}
\label{sec:int}

Graphene -- a single layer of atoms arranged in a two-dimensional (2D) honeycomb lattice --
is a remarkable object in physics for many reasons \cite{Nov04,Nov05}.
It is a transparent and flexible conductor that holds great promise for various material applications \cite{appl1,appl}.
Graphene also displays a number of unusual physical properties which keeps it in focus of
present fundamental research \cite{Neto}.
In particular, the role of saddle point in engineering material properties
has been recently raised in Refs.\ \cite{tG1,tG2}.
There it has been shown that in twisted graphene layers saddle point singularities
seen as two pronounced peaks in the density of electron states (Van Hove singularities \cite{VH})
can be brought very close to the Fermi energy thereby changing their electron characteristics.

In this paper based on the tight binding model on a honeycomb lattice \cite{Ram}
we study magnetic properties of the saddle point of graphene,
namely peculiarities of its Landau levels \cite{Lan0,Lan},
taking into account Harper's broadening \cite{Harp,Wilk} of Landau levels in the presence of a uniform magnetic field $H$.
The solution to the problem is a result of coexistence of two different periods.
The first is given by the electron band structure of graphene and the second is imposed
by the external magnetic field $H$ characterized by its rational flux value,
\begin{eqnarray}
   f = \frac{\phi}{\phi_0} = \frac{p}{q} ,
\label{i1}
\end{eqnarray}
where $\phi$ is the flux through one primitive unit cell, $\phi_0=2\pi \hbar c /e$ and $p$ and $q$ are coprime integers.
($\hbar$ is the reduced Planck constant, $e$ is the electron charge and $c$ is the speed of light.)
Applying the magnetic field to the tight-binding model results in a Hofstadter spectrum \cite{Hof,Ram}.

We however will be more interested in relatively weak (in comparison with its $\pi-$ bandwidth) magnetic fields $H$
with the flux $f=1/q$ where $q$ is large ($q \sim 1000-2000$)
corresponding to $H \sim 40-80$~T, which in principle can be achieved in modern experiments.
Common believe is that at such magnetic fields the system is well described by the semiclassical approach,
where the position of Landau levels is found with the quantization conditions, while the level broadening is negligibly small.
In Ref.\ \onlinecite{Wilk} Wilkinson argued that the Landau levels broadening ($\triangle E$) due to the tunneling between degenerate states has exponential
character, whereas for small $H$ Gao and Niu \cite{GN} estimated it as $\exp(-H_0/H)$, where $H_0$ is a constant.
Thus at $H \rightarrow 0$ ($q \rightarrow \infty$) one obtains a zero width of Landau levels ($\triangle E \rightarrow 0$).
However, the problem is that weakening $H$ leads also to a decrease of the energy separation between two
consecutive Landau levels ($E_0 = \hbar \omega$, where $\omega$ is the cyclotron frequency),
and it makes more sense to speak about the behavior of the ratio $\triangle E / \hbar \omega$.
Surprisingly, while for the vast majority of Landau levels indeed $\triangle E / \hbar \omega \rightarrow 0$ with $H \rightarrow 0$ \cite{Cla},
it does not hold for Landau levels in the immediate vicinity of the saddle point with the energy $\epsilon_M = -t$.
(Here $-t$ is the transfer integral of the tight binding model \cite{Ram}.)
It is this effect that will be of our main concern in the present study.

Recently, Gao and Niu \cite{GN} have proposed a generalized semiclassical quantization condition for closed orbits which
in addition to density of states include response functions in the magnetic field.
It goes beyond the Onsager relation \cite{Ons,Lif2} and is called Roth-Gao-Niu quantization rule in Ref.\ \cite{Fuch}.
Fuchs et al.\ \cite{Fuch} have concluded that the generalized quantization rule is a powerful tool but
it has some limitations.
In particular, as admitted in Ref.\ \cite{GN} the semiclassical theory breaks down near the saddle point.
This also motivates us to consider the saddle point in more detail.

Interestingly, in early work of Azbel \cite{Azb} and Roth \cite{Roth66} a saddle point singularity was already noticed and studied
in the framework of the semiclassical treatment.
At that time though their consideration did not include the saddle point $M$ in graphene but rather related to
a so called self-intersecting open orbit.
In Ref.\ \cite{Roth66} Roth further argued that the self-intersecting open orbit can possibly result in additional
contributions to the magnetic susceptibility including steady and oscillatory parts.
This possibility is concerned in our studying the saddle point singularity.
In contrast to \cite{Azb, Roth66} our analysis will be based on the exact treatment within the tight binding model.

Finally, it is worth mentioning that saddle points exist in all two and three dimensional (3D) electron band structures \cite{VH}.
They often appear on the bordering face of the Brillouin zone (BZ) in the place where a constant energy surface $\epsilon(\vec{q})=\epsilon_0$ touches it \cite{Gla,Mis}.
Therefore, the magnetic effects discussed in the present study are of wide general interest.

The paper is organized as follows.
In Sec.~\ref{sec:sp} we give a brief introduction to the saddle point singularity in graphene.
In Sec.~\ref{sec:res} we present our results for Landau levels in the vicinity of the saddle point energy
obtained by numerical calculations within the tight binding model.
In Sec.~\ref{sec:conc} we summarize our findings and discuss how these effects can be observed experimentally.

\section{Saddle Point Singularity}
\label{sec:sp}

\subsection{Saddle Point in graphene}
\label{sub:sa1}

\begin{figure}[t]\center
\begin{tabular}{l c r}
\includegraphics[width=40mm]{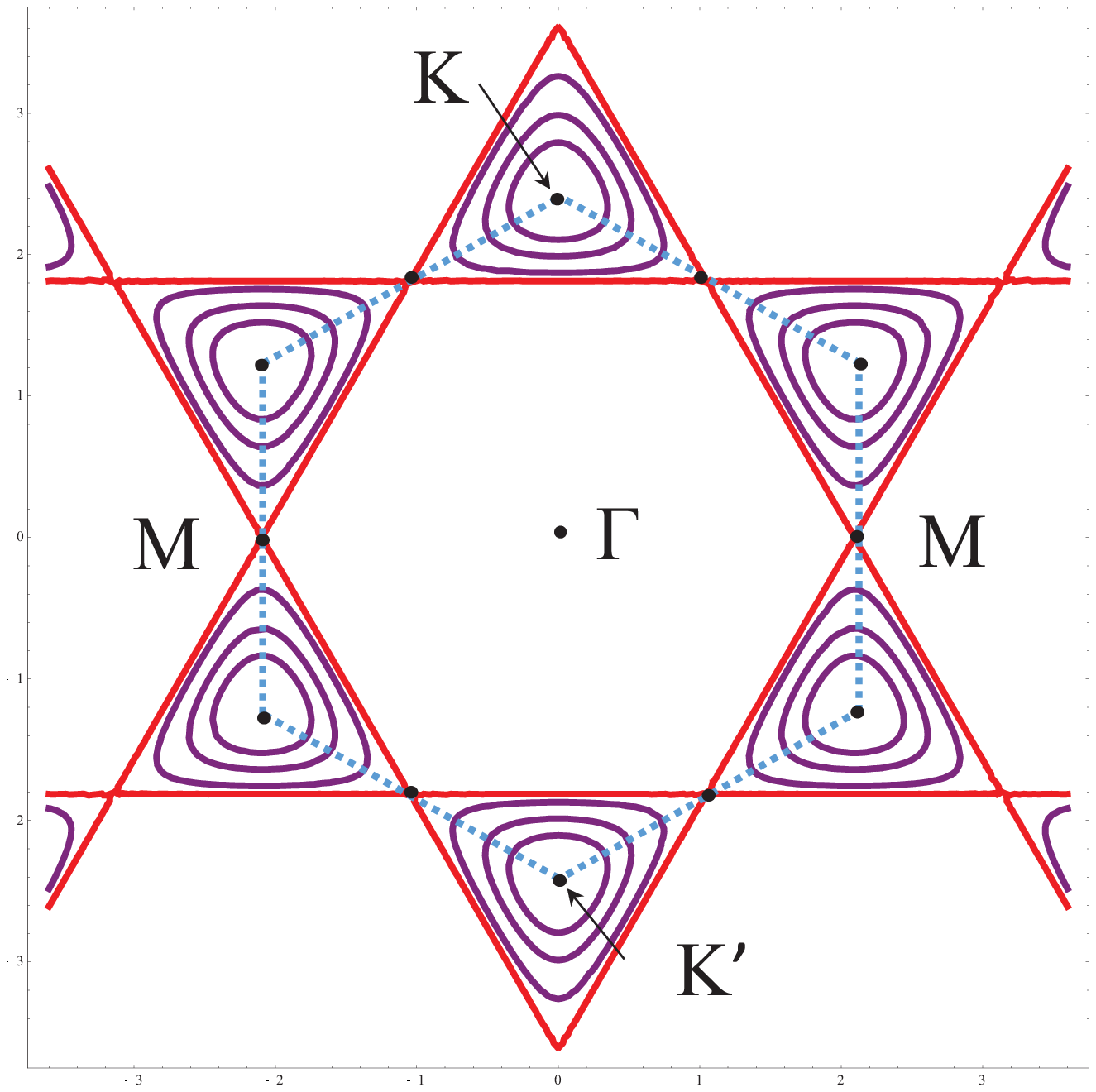}
& \hspace{2mm} &
\includegraphics[width=40mm]{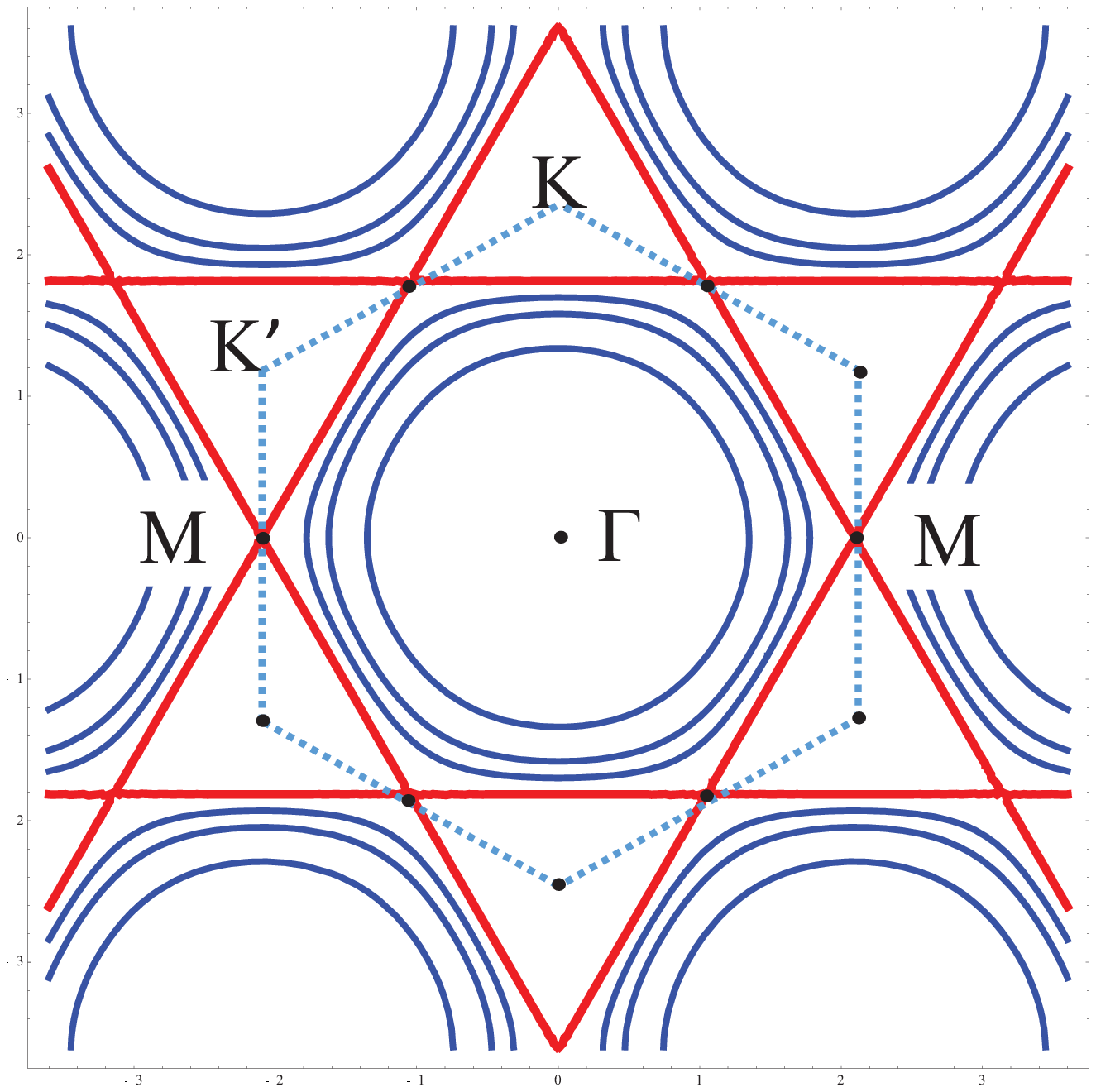}
\end{tabular}
\caption{The first Brillouin zone and two types of semiclassical orbits $O$ of Landau levels in the momentum space of graphene.
Left panel: orbits (purple curves) enclosing the $K$ (or $K'$) point (region I, $\gamma \approx 0$).
Right panel: orbits (blue curves) enclosing the $\Gamma$ point (region II, $\gamma \approx 1/2$)
The straight red lines define an open self-intersecting orbit of the energy $\epsilon_M = -t$ which separates the two regions.
The Brillouin zone is indicated by blue dotted lines. }
 \label{fig1}
\end{figure}

The Brillouin zone (BZ) of the undoped graphene is shown in Fig.\ \ref{fig1}.
Formally, its Fermi surface comprises the $K$ and $K'$ points.
[The coordinates of two points lying on the $y-$axis are given by $\vec{q}_{K,\,K'} = ( 0, \pm 4 \pi/3 \sqrt{3} a )$,
where $a$ is the C-C bond length in graphene.]
The zeroth Landau level ($n=0$) is situated exactly at the Fermi energy
$\epsilon_F = 0$ and in the applied magnetic field $H$ it becomes half populated.
The magnetization of graphene in the tight binding model has been studied by many authors \cite{Mac,Shar,Kish,Pie}.
If the number of electrons in graphene is conserved the half occupation of the Landau level at
the Fermi level is independent of $H$ and hence de Haas - van Alphen (dHvA) oscillations are ineffective.
%

\begin{figure}[!]
\resizebox{0.4\textwidth}{!}
{\includegraphics{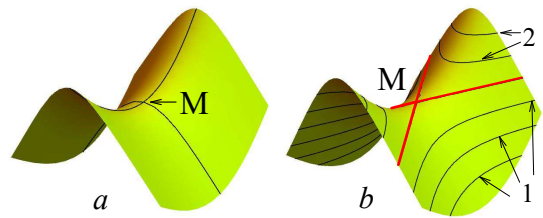}}

\caption{The two dimensional energy profile $E(\vec{q})$ at the saddle point $M$ of graphene.
(a) two characteristic ($xz$ and $yz$) cross-sections; (b) isoenergetic $z-$cross-sections,
giving rise to two types of orbits shown in Fig.~\ref{fig1}: about the $\Gamma$-point (1) and
about the $K$ ($K'$) point (2).
Two straight red lines belong to the open self-intersecting orbit (shown also in Fig.~\ref{fig1}) which separates the two regions.
}
 \label{fig2}
\end{figure}

The saddle points (the $M$ points) are located exactly between two neighboring $K$ and $K'$ points of BZ, Fig.~\ref{fig1}.
In particular, the coordinates of the $M$ point lying on the positive half of the $x-$axis, Fig.~\ref{fig1}, are
$\vec{q}_{M_{+}} = (2 \pi/3 a, 0 )$, whereas
the band energy in the vicinity of the $M \equiv M_+$ point is given by
\begin{eqnarray}
   \epsilon(\vec{q}_{M} + \delta \vec{q}) = -t \left( 1 - \frac{\delta q_x^2}{2 m_x} + \frac{\delta q_y^2}{2m_y} \right)   ,
\label{sp3}
\end{eqnarray}
where $m_x = 2/3$ and $m_y = 2/9$.
Therefore, the $M$ point is a saddle point of the Brillouin zone, whose characteristic two dimensional energy profile is shown in Fig.\ \ref{fig2}.
Since $\vec{\nabla}_q \epsilon(\vec{q}_{M})=0$,
we have the van Hove singularity in the density of states at $\epsilon_M \equiv \epsilon(\vec{q}_{M})$ \cite{VH}.
%
%

\subsection{Saddle Point singularity in the semiclassical treatment}
\label{sub:sa2}

In the semiclassical picture of Landau levels one distinguishes two types of electron orbits in the momentum space
shown on the left and right panels of Fig.\ \ref{fig1}.
In the region I (left panel) with energies $\epsilon_M < \epsilon \leq 0$ the orbits enclose the $K$ and $K'$ points separately,
whereas in the region II (right panel) with energies $-3t < \epsilon < \epsilon_M$ the trajectories
enclose the $\Gamma-$point.

In the momentum space the area $A_n$
enclosed by the $n$th Landau orbit $O_n$ is quantized \cite{Ons,Lif2,Sho,GN,Fuch} according to
\begin{eqnarray}
 A_n =  \frac{2\pi eH}{c \hbar } \left(n + \gamma \right) .
\label{sp4}
\end{eqnarray}
In Eq.\ (\ref{sp4}) $\gamma$ is close to zero for the orbits about the $K$ point (region I) and close to 1/2 for the orbits
about the $\Gamma$ point (region II).
This difference in $\gamma$ is highly nontrivial \cite{GN,Fuch,Xia}.
The zero value of $\gamma$ in graphene at $\epsilon_F=0$ is a manifestation of Berry's phase \cite{MS04}.

\begin{figure}[!]
\resizebox{0.45\textwidth}{!}
{\includegraphics{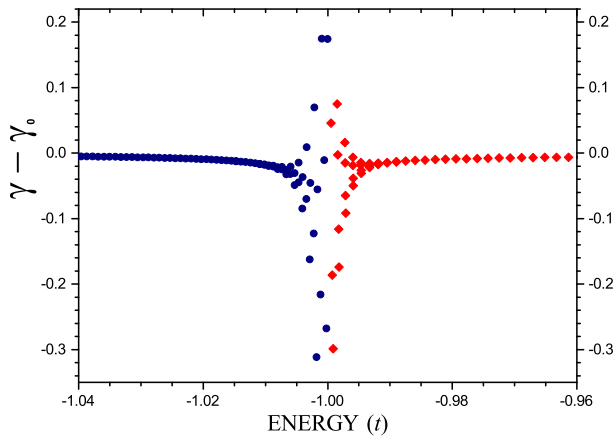}}

\caption{Calculated deviation of $\gamma$, Eq.\ (\ref{sp4}), from $\gamma_0$, where
$\gamma_0=0$ for the orbits encircling the $K$ point (I) and $\gamma_0=1/2$ for the orbits
encircling the $\Gamma$ point (II), see also Fig.~1, main text.
Each point corresponds to an individual Landau level: red circles -- to the region I, blue circles -- to the region II.
Calculations within the tight-binding model \cite{Ram} with $q=1501$, $p=1$ ($H=52.6$~T) and the magnetic wave number $k=0$.
}
 \label{fig1s}
\end{figure}

The exact relations $\gamma_0 = 0$ and $\gamma_0 = 1/2$ hold only for smallest orbits at the point $K$ and $\Gamma$ of BZ.
For other orbits passing through other points of BZ $\gamma$ can differ from these values.
Our calculations of $\gamma$ for other Landau levels within the tight binding model (with $k_x=k_y=0$), shown in Fig.~\ref{fig1s},
demonstrates that the deviations of $\gamma$ from $\gamma_0 = 0$ in the region I and from $\gamma_0 = 1/2$ in the region II are very small
and become noticeable and large only for orbits in a very narrow energy region around $\varepsilon_M = -t$, i.e.
in the immediate neighborhood of the saddle point $M$.
%
%
Small deviations of $\gamma$ (except at the $M-$point) are in agreement with estimations of the effect of
non-parabolicity and band curvature made in Ref.\ \cite{For}.
Scattered data of $\gamma$ at $\varepsilon_M$, on the other hand, is an indication
that the saddle point is very different from all other points of the Brillouin zone.

A special role of the saddle point can be easily understood in the semiclassical theory.
A singular bordering orbit $O_M$ at $\epsilon = \epsilon_M$
shown by straight red lines in Figs.\ \ref{fig1} and \ref{fig2}, separates BZ in regions I and II and
represents a so-called self-intersecting open orbit.
(Earlier, a saddle point has been considered as a part of a hypothetical `figure eight' self-intersecting orbit by Azbel \cite{Azb} and Roth \cite{Roth66}.)
In contrast to the other trajectories, the movement of an electron along the self-intersecting open orbit is not limited to a certain region in the momentum space
and therefore it should have a certain dispersion relation even in a weak magnetic field.
In next section we will see that Landau levels in the neighborhood of the saddle point are broadened
in magnetic minibands whose band width is comparable with the energy difference between two Landau levels
even in weak magnetic fields.

\subsection{Tight Binding Model}
\label{sub:tb}

In the following we work within the tight-binding model of Ref.\ \onlinecite{Ram}, whose Hamiltonian
reads as
\begin{eqnarray}
   {\cal H} = -t \sum_{i,j} \exp(i \theta_{ij})\, c_i^{\dagger} c_j ,
\label{sp1}
\end{eqnarray}
where $c_i^{\dagger}$, $c_j$ are the electron creation and annihilation operators defined on the honeycomb lattice
sites $i$, $j$ of graphene (with coordinates $\vec{r}_i$, $\vec{r}_j$);
$-t$ ($t>0$) is the transfer integral between two neighboring sites,
and
\begin{eqnarray}
   \theta_{ij} = \frac{2 \pi}{\phi_0} \int_{r_i}^{r_j} \vec{A} \, d \vec{\ell} .
\label{sp1}
\end{eqnarray}
Here $\vec{A}$ is the vector potential defined by the magnetic field $H$, and $d \vec{\ell}$ is an element of the path (straight line)
connecting the sites $i$ and $j$.
This task for magnetic fields with the rational flux $f = p/q$, is reduced to a system of one dimensional
finite difference equations (along the $x-$axis) depending parametrically on $k_y$ \cite{Ram}.
Introducing Bloch basis functions defined by $k_x$ leads then to an eigenspectrum problem \cite{Wilk}.
Therefore, finally one deals with the $2q \times 2q$ complex Hermitian eigenvalue problem which is solved numerically for every point $(k_x,k_y)$ belonging
to the magnetic Brillouin zone (see Fig.~3, main text).
Interestingly, as shown in Appendix \ref{sec:app} in the case of odd $p$ and $q$ and $k_x=k_y=0$ the spectrum can be obtained by solving
a $q \times q$ real symmetrical eigenvalue problem.

In the present paper we are interested in relatively weak magnetic fields ($H \sim 40-80$~T)
with the flux $f=1/q$ where $q$ is a large integer ($q \sim 1000-2000$), when most of Landau minibands are very narrow ($< 10^{-8}\, t$).
In this paper for simplicity we have neglected
the Zeeman electron energy due to different spin polarization, which however can be easily introduced in the end.

\section{RESULTS}
\label{sec:res}

\subsection{Broad Landau minibands}
\label{sec:wide}

\begin{figure}[!]
\resizebox{0.45\textwidth}{!}
{\includegraphics{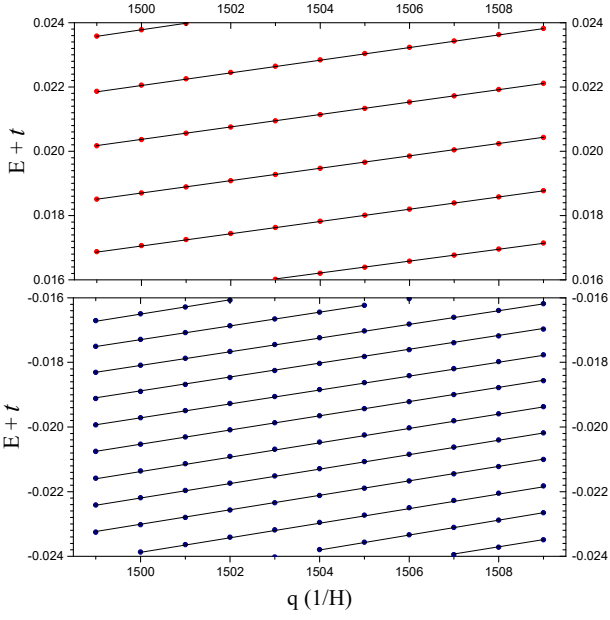}}

\caption{Typical energies of Landau levels (in units of $t$) and their change with $1/H=q$ ($p=1$) above (red circles) and below (blue circles)
the saddle point energy ($\epsilon_M + t = 0$)
calculated within the tight binding model, Ref.\ \onlinecite{Ram}.
($H$ changes from 52.7 T to 52.3 T.)
The width of all shown Landau levels is practically zero ($<10^{-8}\, t$).
}
 \label{fig2s}
\end{figure}

We start by considering the Landau levels above and below the saddle point energy $\epsilon_M=-t$,
excluding a relatively thin energy region at its immediate vicinity,
from $\epsilon_M-\delta$ to $\epsilon_M+\delta$, where $\delta = 0.0164\; t$.
Solving numerically the equations of Ref.\ \cite{Ram,Wilk} for $q=1499-1509$ ($p=1$) corresponding
to the magnetic field values 52.3--52.7~T, we find that the width of these Landau levels is very narrow.
Energies of some Landau levels are shown in Fig.\ \ref{fig2s}.
%
%
It is less than $6\times 10^{-9}\; t$ for the lower part and $4\times 10^{-9}\; t$ for the upper part of the spectrum.
If such a behavior had persisted down to the saddle point, it would have resulted in oscillations of the magnetic susceptibility
of the dHvA type due to the abrupt appearance and disappearance of Landau levels at $\epsilon_M$, Ref.\ \onlinecite{Roth66}.
It is also worth noting that the energy position of these Landau levels is in very good correspondence
with the values obtained with the quantization condition, Eq.\ (\ref{sp4}).
If we use $\gamma=0$ in the Region I and $\gamma=1/2$ in the Region II the maximal energy mismatch is only $2.6\times 10^{-6}\; t$
and $1.6\times 10^{-5}\; t$, respectively.
On the other hand, if we reverse Eq.\ (\ref{sp4}) by fitting $\gamma$ to the calculated energy values we obtain
that the maximal deviation of $\gamma$ is $1\times 10^{-2}$.
In calculating the energy properties of these Landau levels we can safely take $k_x=k_y=0$ for the magnetic wave number \cite{Ram}.

However, the situation for the saddle point region, from $\epsilon_M-\delta$ to $\epsilon_M+\delta$, is very different from the picture discussed above.
A clear manifestation of this fact are large oscillatory deviations of $\gamma$ from their theoretical values, illustrated in Fig.~\ref{fig1s}.
Our calculations within the tight binding model yield
that each Landau level is rather a magnetic miniband characterised by the dispersion law $E_m(\vec{k})$, where $\vec{k} \equiv (k_x,k_y)$.
One finds that
\begin{eqnarray}
 E_m(k_x,k_y) = E_m(-k_x,k_y) = E_m(k_x,-k_y) \nonumber \\
 = E_m(-k_x,-k_y) .
\label{w1}
\end{eqnarray}

\begin{figure}[!]
\resizebox{0.4\textwidth}{!}
{\includegraphics{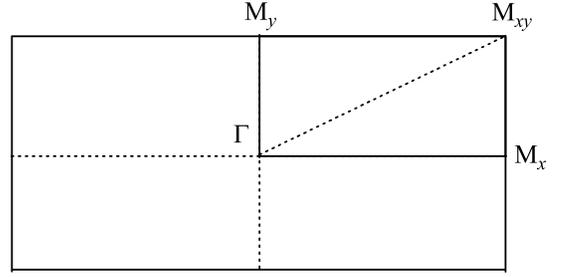}}

\caption{Magnetic Brillouin zone of the honeycomb lattice and its irreducible part ($\Gamma-M'_x-M'_{xy}-M'_y-\Gamma$).
Here $M'_x(\frac{\pi}{2qd_x},0)$, $M'_y(0,\frac{\pi}{2qd_y})$, $M'_{xy}(\frac{\pi}{2qd_x},\frac{\pi}{2qd_y})$,
where $d_x=3a/2$ and $d_y=a\sqrt{3}/2$ ($a$ is the C-C bond length in graphene).
}
 \label{fig3}
\end{figure}

We now have to define the magnetic Brillouin zone (or the magnetic primitive unit cell).
In real space the difference equations along the $x-$axis [i.e. Eq.\ (4.6) of Ref.\ \cite{Ram}] are periodic with the shortest translation vector $2q d_x$,
where $d_x=3a/2$ ($a$ is the C-C bond length in graphene).
This implies the shortest translation vector $2\pi/2 q d_x =\pi/q\, d_x$ along the $k_x$-direction in the reciprocal space.
Analogously, one can show that the difference equations \cite{Ram} have the shortest period $\pi /q\, d_y$ along the $k_y$-direction,
where $d_y=a\sqrt{3}/2$.
The primitive magnetic unit cell then comprises a very small region in $k$-space (especially when $q \rightarrow \infty$):
$-\pi/2qd_x \leq k_x \leq \pi/2qd_x$ and $-\pi/2qd_y \leq k_y \leq \pi/2qd_y$.
Taking into account Eq.\ (\ref{w1}), we introduce corner points of this rectangular primitive unit cell: $M'_x(\pi/2qd_x,0)$, $M'_y(0,\pi/2qd_y)$ and $M'_{xy}(\pi/2qd_x,\pi/2qd_y)$, Fig.\ \ref{fig3}.
Thus, the irreducible part of the magnetic Brillouin zone is fully represented by its forth part given by the
rectangle $\Gamma-M'_x-M'_{xy}-M'_y$. In the irreducible part of the magnetic BZ we have defined a 45$\times$45 mesh (2025 points)
which has been used for calculation of the magnetic band widths.
[One can also plot the dispersion dependencies along the high symmetry lines $\Gamma-M'_y-M'_{xy}-\Gamma-M'_x-M'_{xy}$,
shown below in Fig.\ \ref{fig6} and \ref{fig8}.]
%
%

\begin{figure}[!]
\resizebox{0.48\textwidth}{!}
{\includegraphics{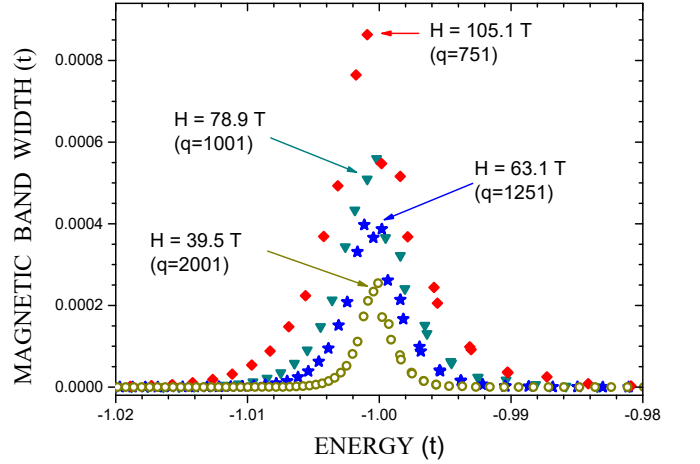}}

\caption{Calculated widths of magnetic minibands in the neighborhood of the saddle point $M$ ($\epsilon_M=-t$).
Each point represents an individual magnetic band with the $x-$coordinate corresponding to the middle band energy.
}
 \label{fig4}
\end{figure}

\begin{figure}[!]
\resizebox{0.48\textwidth}{!}
{\includegraphics{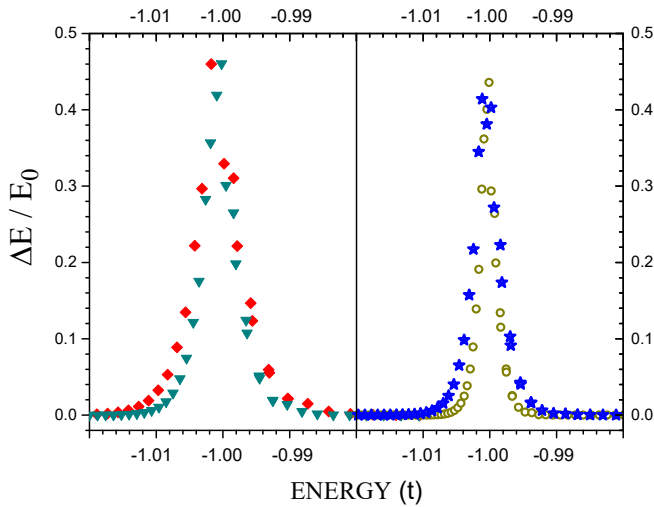}}

\caption{Widths of magnetic minibands in the neighborhood of the saddle point $M$ ($\epsilon_M=-t$)
in respect to an average energy difference $E_0=\hbar \omega_0$.
Each point represents an individual magnetic band. Notations are the same as in Fig.\ \ref{fig4}: red diamonds -- $H=105.1$~T ($q=751$),
green triangles -- $H=78.9$~T ($q=1001$), blue stars -- $H=63.1$~T ($q=1251$), hollow dark yellow circles -- $H=39.5$~T ($q=2001$).
}
 \label{fig5}
\end{figure}

Our results for various magnetic fields are represented in Fig.\ \ref{fig4}.
%
%
Inspection of the figure reveals that the saddle point $M$ is a singularity resulting in a substantial band width of
all Landau levels lying in its vicinity.
However, the band width peaks centered at $\epsilon_M$
become narrower and smaller
with weakening the magnetic field $H$ (with increasing $q$).
One might be tempted to conclude that in the limit of $H \rightarrow 0$ ($q \rightarrow \infty$) it is possible to suppress the level broadening
to zero and eventually to get rid of the effect altogether, but this is incorrect.
The problem is that the energy difference between magnetic bands, $E_0=\hbar \omega$
(where $\omega$ is the cyclotron frequency)
also decreases with $H$, and the question is whether
at $H \rightarrow 0$ the band widths are reduced in respect to $E_0$.
Our calculations show that this is not the case, Fig.\ \ref{fig5}.
%
%
In Fig.\ \ref{fig5} we plot the ratio of the $n-$th bandwidth $\triangle E(n)$ to an average energy value $E_0=\hbar \omega_0$ between
neighboring minibands. ($E_0$ is averaged over an energy range of $\delta = 0.03\, t$ below $\epsilon_M$
excluding a few wide bands at $\epsilon_M$.)
From Fig.\ \ref{fig5} it follows that the largest ratios reach the value of $0.4-0.5$ meaning that the
band span is very substantial on the scale of $\hbar \omega$.
Calculated parameters of the Landau miniband with the maximal bandwidth for various $H$ are quoted in Table \ref{tab1}.
An inspection of the Table reveals that there is no tendency of decreasing $\triangle E / \hbar \omega$ with weakening $H$ and
the situation is likely to persist for
even smaller magnetic fields ($H < 39.5$~T and $q > 2001$).
%
\begin{table}
\caption{Maximal band width $\triangle E$ ($\triangle E^*$) in the immediate neighborhood of the saddle point $M$ with the energy $\epsilon_M=-t$.
$E_c$ is the band energy center, $E_0=\hbar \omega_0$ is an averaged energy between two subsequent Landau levels.
Here $H$ is in Tesla, while $E_c + \epsilon_M$, $\triangle E$ and $E_0$ are in units of $10^{-3}\, t$. $\triangle E^*$ is in meV
assuming $t=2.8$~eV \cite{Neto}.
\label{tab1} }

\begin{ruledtabular}
\begin{tabular}{l  c  c  c  c  c  c }

  $q$ & $H$ &  $E_c + \epsilon_M$ &  $\triangle E$  & $\triangle E^*$ (meV) & $E_0$ & $\triangle E / E_0$ \\
\tableline
 751  &   105.1  & -0.907 & 0.863 & 2.42 & 1.662 & 0.519 \\
 1001 &    78.9  & -0.202 & 0.560 & 1.57 & 1.215 & 0.460 \\
 1251 &    63.1  & -1.131 & 0.398 & 1.11 & 0.960 & 0.415 \\
 1502 &    52.6  & -0.551 & 0.398 & 1.11 & 0.698 & 0.571 \\
 1504 &    52.5  & -0.269 & 0.368 & 1.03 & 0.697 & 0.528 \\
 1800 &    43.8  & -0.103 & 0.286 & 0.80 & 0.652 & 0.439 \\
 2001 &    39.5  & -0.091 & 0.255 & 0.71 & 0.583 & 0.436 \\

\end{tabular}
\end{ruledtabular}
\end{table}

\subsection{Dispersion law and density of states}
\label{sec:bands}

\begin{figure}[!]
\resizebox{0.48\textwidth}{!}
{\includegraphics{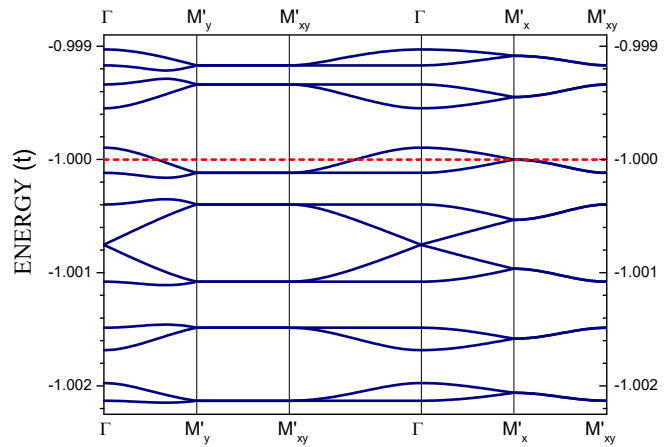}}

\caption{Calculated magnetic band dispersions in the immediate vicinity of the saddle point, $H=52.6$~T ($q=1502$, $p=1$).
Each point corresponds to an individual $k-$value in the irreducible part of the magnetic BZ, Fig.\ \ref{fig3}. The total number of $k-$points is 4000.
The saddle point energy ($\epsilon_M = -t$) is shown by the red line.
}
 \label{fig6}
\end{figure}

\begin{figure}[!]
\resizebox{0.48\textwidth}{!}
{\includegraphics{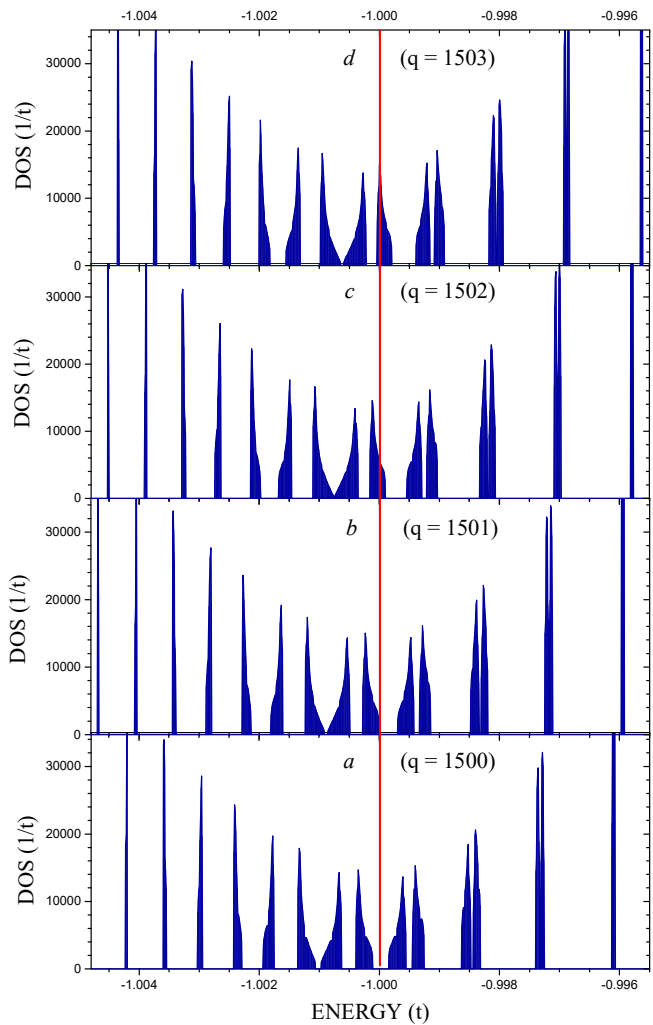}}

\caption{Calculated density of states of Landau minibands in the neighborhood of the saddle point in the external magnetic field $H$:
(a) $H=52.63$~T ($q=1500$); (b) $H=52.59$~T ($q=1501$); (c) $H=52.56$~T ($q=1502$); (d) $H=52.52$~T ($q=1503$).
In all cases $p=1$, the $45 \times 45$ $k-$mesh (2025 points) has been used.
The red line indicates the energy at the saddle point.
}
 \label{fig7}
\end{figure}

For calculations of the dispersion law and the density of states of minibands in the immediate
neighborhood of the saddle point
we have used a set of 200 $k$-points along high symmetry lines of the irreducible part
of the Brillouin zone (Fig.\ \ref{fig3}), and a $45 \times 45$ $k-$mesh (2025 points), respectively.
In Figs.\ \ref{fig6} and \ref{fig7}c we reproduce our results for $q=1502$ ($p=1$) corresponding to the magnetic field $H=52.6$~T.
%
%
The picture of density of states shown in Fig.\ \ref{fig7}c is obtained by replacing the delta-function at each $E(k)$ by
the gaussian function with $\sigma=9 \times 10^{-6}\; t$.

Fig.\ \ref{fig6} clearly confirms the fact that the band width is comparable with the energy separation between neighboring bands
discussed earlier in Sec.\ \ref{sec:wide}, although the chosen value of $q=1502$ is relatively large and the corresponding magnetic
field is relatively weak.

\begin{figure}[!]
\resizebox{0.48\textwidth}{!}
{\includegraphics{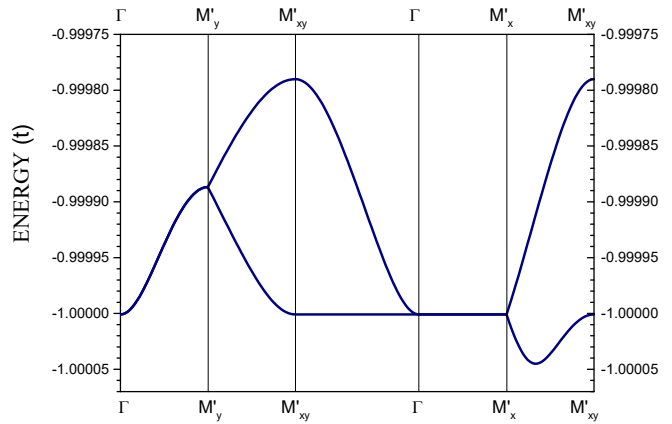}}

\caption{Calculated dispersion law of the saddle point magnetic band for $q=1503$, $p=1$ ($H=52.52$~T).
Each point corresponds to an individual $k-$value in the irreducible part of the magnetic BZ, Fig.\ \ref{fig3}.
The total number of $k-$points is 4000.
}
 \label{fig8}
\end{figure}

Interestingly, in Figs.\ \ref{fig6} and \ref{fig8} one finds a band around the energy $\epsilon_M$ (red line in Fig.\ \ref{fig6}) whose energy values can lie
above and below $\epsilon_M$. In terms of the semiclassical language it contains two types of different orbits -- those
enclosing the $K$ and $\Gamma$ points and belonging to different regions of the Brillouin zone -- I and II, Fig.\ \ref{fig1} and Sec.\ \ref{sub:sa2}.
For the whole $\Gamma-M_x$ branch of the band in Fig.\ \ref{fig8} ($q=1503$) one finds
that its energy is very close to $\epsilon_M=-t$, corresponding to
the open self-intersecting orbit shown by red in Fig.\ \ref{fig1}.

We next investigate the dependence of the magnetic band structure on the change of the magnetic field $H$ and the flux number $q$.
The calculated spectra for $q$ from 1500 to 1503 are given in Fig.\ \ref{fig7}.
%
%
We observe that on increasing $q$ and decreasing $H$ the bands move upward in energy crossing the saddle point value of $\epsilon_M$
on its way (the red line in Fig.\ \ref{fig7}).
The bands become wider on approaching $\epsilon_M$ and thinner
as they get farther away from it. The largest width belongs to a band with the energy just below $\epsilon_M$.
The saddle point energy $\epsilon_M$ can lie in the band gap as for $q=1500$ and $q=1504$ (not shown) or
be inside a magnetic band as for $q=1501$, 1502, 1503, Fig.\ \ref{fig7}.
Similarly to the behavior of all narrow Landau levels (e.g. Fig.~\ref{fig2s}) the band picture pattern is periodic in $q$ (and $H$).
%
%

\subsection{Consequences for longitudinal electron transport in electric field}
\label{sec:transp}

Here we briefly discuss some consequences of the broad Landau minibands for
electron transport when an external electric field $E$ is applied along the $x$-axis.
The problem of semiclassical electron dynamics in magnetic Bloch bands
has been solved by Chang and Niu in Refs.\ \cite{Chang95,Chang96}.
They considered $E$ and an additional magnetic field $\delta B$
as perturbing fields put on top of the magnetic Bloch states formed in the
presence of a strong magnetic field $B$. $\delta B$ is needed
to account for the irrational total magnetic field $B + \delta B$,
in our case we can assume $\delta B=0$. Then the velocity
for an electron in the state $n$ is given by
\begin{eqnarray}
 \vec{v}_n = \frac{\partial E_m^n(\vec{k})}{\hbar\, \partial \vec{k}} - \frac{e}{\hbar} \vec{E} \times \Omega_n(\vec{k}) ,
\label{t1}
\end{eqnarray}
where $\Omega_n$ is the Berry curvature of the $n$th magnetic band \cite{Xia}.
It can be shown that the second part of Eq.\ (\ref{t1}) represents
an anomalous velocity which being always transverse to the electric field contributes to the Hall conductivity,
whereas the first part is the usual band dispersion term responsible for the longitudinal conductivity \cite{Xia}.

In the following we limit ourselves to the longitudinal part, leaving the more complicated Hall contribution
to future analysis.
Notice that the dispersionless parts of energy spectrum like the $\Gamma-M'_x$ branch shown in Fig.\ \ref{fig8}
and some others, result in zero contribution
to the longitudinal current since there $v_{n,x} = \partial E_m(\vec{k}) / \hbar\, \partial k_x = 0$.
These states can be considered as localized which do not support longitudinal electric current.
As shown in Ref.\ \cite{Chang96} at zero
temperature the longitudinal current can be reduced to the following expression:
\begin{eqnarray}
 J_{xx} = e^2 \tau g(E_F)\, D E ,
\label{t2}
\end{eqnarray}
where $\tau^{-1}$ is the impurity scattering rate, $D$ is the diffusion tensor,
and $g(E_F)$ is the density of states of the magnetic band at the Fermi level
(see details of the definitions in Ref.\ \cite{Chang96}.
(Here we assume that $\partial \mu /\partial \vec{r} = 0$ and thus there is no contribution
due to $J^{\mu}$, Eq.\ (3.6) of Ref.\ \cite{Chang96}).
Although in pristine graphene the Fermi level is situated far above the saddle point
minibands, there are effective ways discussed in Sec.\ \ref{sec:conc} below
to bring $E_F$ close to them (for example, this is experimentally achievable in twisted graphene \cite{tG1,tG2}).
In that case from Eq.\ (\ref{t2}) it follows that the longitudinal conductivity $\sigma_{xx}$
is proportional to the magnetic density of states $g(E)$, taken at $E=E_F$.
Therefore, the results obtained within the present approach, e.g. the calculated
magnetic density of states $g(E)$ reproduced in Fig.\ \ref{fig7}, can be tested
by changing the position of the Fermi level
and measuring the longitudinal conductivity $\sigma_{xx} \propto g(E_F)$
at zero temperature.

\subsection{Contribution to magnetic susceptibility}
\label{sec:sus}

We have also calculated the energy contribution from the saddle point region to the total energy change
in the presence of the external magnetic field at zero temperature $T=0$.
Since in the immediate neighborhood of $\epsilon_M$ the magnetic levels demonstrate dispersion,
for every Landau miniband $n$ we have performed integration
throughout the irreducible part of the magnetic Brillouin zone, Fig.\ \ref{fig3}, to calculate its energy,
\begin{eqnarray}
  E_n = \sum_{k_x, k_y} E_n(k_x, k_y)\, w(k_x, k_y)  ,
\label{s1}
\end{eqnarray}
where $w$ is an effective weight factor at $k_x, k_y$.
A typical displacement $\delta E_n = E_n - E_n^{\Gamma}$ of the averaged band energy $E_n$, Eq.\ (\ref{s1}), from its $\Gamma$ value $E_n^{\Gamma} = E_n(\vec{k}=0)$
is comparable to the band width, with largest energy shifts $\delta E_n$ found at $\epsilon_M$.

Calculated energy changes for states below and above $\epsilon_M$ within the saddle point region
on applying the external magnetic field of $H=78.9$~T ($q=1001$, $p=1$) are quoted in Table \ref{tab2}.
%
\begin{table}
\caption{Contribution to the energy change in the magnetic field $H=78.9$~T ($q=1001$, $p=1$) from
the Landau levels with energies from $E_{dn}$ to $E_{up}$ (including the saddle point), in units of $10^{-3} \times t$.
($E_{dn}$ and $E_{up}$ are in units of $t$).
$\triangle E(I)$ is the contribution from the region I lying above $\epsilon_M$
(i.e. from $\epsilon_M$ to $E_{up}$),
$\triangle E(II)$ is the contribution from the region II lying below $\epsilon_M$ (from $E_{dn}$ to $\epsilon_M$),
$\triangle E = \triangle E(I) + \triangle E(II)$.
\label{tab2} }

\begin{ruledtabular}
\begin{tabular}{l  c  c  c  c  c  c }

  $E_{up}$ & -0.99 & -0.98 &  -0.9 & -0.75  & -0.5 & 0.0 \\
  $E_{dn}$ & -1.01 & -1.02 &  -1.2 & -1.50  & -3.0 & -3.0 \\
\tableline
 $\triangle E(I)$ &  4.343  &  4.090 &  3.892 &  3.848 &  3.826 &  4.493 \\
 $\triangle E(II)$  & -3.408  & -3.579 & -3.779 & -3.798 & -3.836 & -3.836 \\
 $\triangle E$     &  0.935  &  0.511 &  0.113 &  0.050 & -0.009 &  0.657 \\

\end{tabular}
\end{ruledtabular}
\end{table}
In both regions the change of energy was calculated according to
\begin{eqnarray}
  \triangle E(i) = E(i,\,H \neq 0) - E(i,\,H = 0) ,
\label{s2}
\end{eqnarray}
where $i=I,\, II$ and $E(i,\,H \neq 0)$, $E(i,\,H = 0)$ are the corresponding energies in the presence and absence of $H$.
Although the values of $\triangle E(I)$, $\triangle E(II)$ are relatively large even for small widths of the chosen interval (i.e., $E_{up} - E_{dn}$),
they tend to compensate each other, especially on increasing $E_{up} - E_{dn}$. This is clearly shown in Table \ref{tab2} (last row)
for the quantity $\triangle E = \triangle E(I) + \triangle E(II)$.
Last column of Table \ref{tab2} includes all states below the Fermi energy $\epsilon_F = 0$.
The rise of $\triangle E$, which in this case corresponds to the energy change for all occupied electron states in graphene, is
due to the contribution from the zero half-populated Landau level at $\epsilon_F = 0$ \cite{Mac}.
Our estimations give the value of 6.72$\times 10^{-3}\, t$ for diamagnetic energy change from the Fermi energy region
(around the $K-$point). Therefore, the contributions from the other regions of the Brillouin zone amount to
only $-0.015\times 10^{-3}\, t$, i.e. 2{\%} of the total effect.

The absence of oscillations due to the saddle point anomaly contradicts expectations obtained within the semiclassical
scenario \cite{Azb,Roth66}. It can be understood in a simplified way as following.
In the semiclassical picture Landau levels situated in the region I and II, Fig.\ \ref{fig1}, are completely uncorrelated.
They can approach the open self-intersecting orbit and then disappear independently, similarly
to what happens with Landau levels passing through the Fermi surface. In the quantum case a Landau level just below $\epsilon_M$
is connected with a close Landau level with energy above $\epsilon_M$.
This relation is clearly shown in Fig.~\ref{fig7}, where the pattern of Landau minibands is approximately conserved in shape and
simply shifts upward as a whole on decreasing $H$.
As discussed in Ref.\ \cite{Nik1} all fully occupied Landau levels situated below the Fermi energy do not contribute to the diamagnetic effect.

\section{Discussion and Conclusions}
\label{sec:conc}

In conclusion, based on the tight binding calculation of the single layer graphene in the external magnetic field $H$ with the rational flux $f = 1/q$
we have studied the saddle point ($M$) anomaly of Landau levels.
We find that at the saddle point energy $\epsilon_M$ the Landau levels become broadened into bands reaching the maximal width of 0.4-0.5
of the energy separation
between two levels ($\hbar \omega$) even in relatively small magnetic field, $H \sim 39.5$~T ($q=2001$, $p=1$), Fig.\ \ref{fig5}.
These Landau levels or magnetic minibands cross the saddle point energy $\epsilon_M$ on increasing $q$ (or decreasing $H$), Fig.\ \ref{fig7}.
The energy $\epsilon_M$ can be located in a band gap or inside a band.
For $q=1502$ and $q=1503$ ($p=1$) the magnetic dispersion laws $E(k_x,k_y)$ are reproduced in Figs.\ \ref{fig6} and \ref{fig8}, respectively. 

In the magnetic field with rational flux $f=p/q$ the saddle point structure of minibands does not affect the magnetic response of graphene in the magnetic field,
which is dominated by the contribution from electron states at the Fermi level.
This is in line with the general conclusion that the diamagnetic response for 3D metals is caused by a narrow region near the Fermi level \cite{Nik1}.
Therefore, a statement about possible dHvA oscillations due to the irregular character of Landau levels at $\epsilon_M$
made on the basis of semiclassical picture \cite{Roth66} is not confirmed by our calculations.
In contrast to the semiclassical theory considering Landau levels in region I and II as being completely independent,
the tight binding model results in  a transitional region with
broad minibands at $\epsilon_M$, in which the structure of Landau levels is essentially conserved and shifts as a whole on
changing $H$.

Although the broad Landau minibands discussed in this paper appear even in small magnetic fields their energy ($\epsilon_M$)
lies well below the Fermi energy $\epsilon_F$.
Therefore, for observing the peculiarities of the broad magnetic bands explicitly, the doping of graphene should be very strong.
One way to approach the saddle point $\epsilon^*_M = +t$ in the virtual $\pi^*$-energy band is the intercalation of graphene with alkali
or alkaline earth metals \cite{Petr,intercal}.
Also, one can effectively change the Fermi level in graphene in various graphene heterostructures
where parameters of the band structure can be tuned to designed values \cite{hetero1,hetero2} or by substituting carbon atoms with boron
or nitrogen \cite{subst}. The singularity can also be investigated in `artificial graphene' or, more precisely, in
artificially prepared hexagonal lattices which provide regimes of parameters not accessible in natural graphene \cite{artG}.

However, the most promising experiment is offered by twisted graphene layers' setup \cite{tG1,tG2}.
There, low energy saddle points singularities in twisted graphene layers are observed as two Van Hove peaks in the
density of states measured by scanning tunnelling spectroscopy.
Moreover, these singularities can be brought arbitrary close to the Fermi energy by varying the angle of rotation \cite{tG1,tG2}.
Therefore, using the twisted graphene layers and gating technique in an applied magnetic field one can
study the crossing of the Fermi energy by a broad Landau level considered in this paper.

If the Fermi energy is brought very close to the Van Hove peaks (saddle points) there will be
experimental manifestations of accompanied broad Landau states in the (half-integer) quantum Hall effect (QHE),
Shubnikov-de Haas oscillations (SdHOs) and other transport properties.
While all these effects require a separate scrupulous analysis some predictions can be made at the present stage
of investigation. In particular, based on the electron dynamics in magnetic Bloch bands developed in Refs.\ \cite{Chang95,Chang96},
we conclude that at low temperatures the longitudinal conductivity $\sigma_{xx}$ is proportional to the magnetic density of states (MDOS), Eq.\ (\ref{t2}),
and therefore, in principle one can restore the calculated MDOS (e.g. shown in Fig.\ \ref{fig7})
by changing $E_F$ within a miniband and measuring the correspondent $\sigma_{xx}$.

These considerations refer to the pristine 2D graphene. In the case of narrow graphene nanoribbons one should take into account
the various types of edges which can exist in the graphene layer \cite{edge,Neto}.
In particular, zigzag edges sustain surface electron states at $\epsilon_F=0$ localized on the edges \cite{edge,Neto}. In the single layer graphene these states are well
separated from the Van Hove singularities and are not expected to influence transport properties of broad Landau levels.

Finally, we remark that the existence of saddle points is a topological effect \cite{VH}.
In general, in the 2D case there must be at least two saddle points for each band energy branch \cite{VH} and the only problem is how far
they are from the Fermi energy. Any saddle point has the self-intersecting orbit causing the broadening of nearest Landau levels.
Therefore, the effect should also occur in the square lattice and in all other
2D materials.
The same applies to the 3D structures, although in this case the saddle points should be defined in respect to
the energy dependence in planes in the $q$-space which are perpendicular to the direction of $H$.

\acknowledgements
The author acknowledges helpful discussions with A. V. Rozhkov.

\appendix

\section{Reduction to the $q \times q$ eigenproblem}
\label{sec:app}

Here we show that the spectrum of the tight binding model Ref.\ \onlinecite{Ram} for odd $q$ and $p$ at the
magnetic $\Gamma-$point ($k_x = k_y = 0$)
can be obtained by the diagonalization of a $q \times q$ real Hamiltonian matrix $H^{red}$ whereas for other cases
it requires the diagonalization of the full $2q \times 2q$ complex Hermitian matrix $H^{full}$.
The full spectrum then is obtained by doubling the eigenspectrum of $H^{red}$.
This observation makes much easier calculations of energy spectrum for that particular case.

The matrix elements of $H^{full}$ at $\Gamma$ are given by \cite{Ram,Wilk}
\begin{eqnarray}
  & & H^{full}_{m,m} = 2 \cos\left[ -2 \pi \frac{p}{q} \left( m + \frac{1}{2} \right)  \right],     \label{a1a} \\
  & & H^{full}_{m,m+1} = 2 \cos\left[ - \pi \frac{p}{q} \left( m + \frac{1}{2} \right)  \right],     \label{a1b} \\
  & & H^{full}_{m+1,m} = H^{full}_{m,m+1}.  \label{a1c}
\end{eqnarray}
Here $t = 1$. One can show that
\begin{eqnarray}
  & & H^{full}_{m,m} = H^{full}_{m+q,\, m+q},     \label{a2a} \\
  & & H^{full}_{m+1,m} = -H^{full}_{m+q,\, m+q+1}.  \label{a2b}
\end{eqnarray}
If $q$ is odd, it can be written as $q = 2l + 1$ ($l < q$) and then for odd $p$ we arrive at
\begin{eqnarray}
  & & H^{full}_{l,l} = -2,     \label{a3a} \\
  & & H^{full}_{l,\, l+1} = H^{full}_{l+1,\, l} = 0.  \label{a3b}
\end{eqnarray}
From Eqs.\ (\ref{a2a}), (\ref{a2b}) it follows that the same relations hold for $l+q$.
This implies that by changing order of basis functions the matrix $H^{full}$
can be transformed to the block diagonal form,
\begin{eqnarray}
 H^{full} =  \left[ \begin{array}{c c}
            H' & 0 \\
            0 & H'' \\
            \end{array}  \right],
 \label{a4}
\end{eqnarray}
where both $H'$ and $H''$ are $q \times q$ matrices, which now can be diagonalized separately.
Further, from Eqs.\ (\ref{a2a}), (\ref{a2b}) it follows that
\begin{eqnarray}
  & & H'_{m,m} = H''_{m+q,\, m+q},     \label{a5a} \\
  & & H'_{m+1,m} = -H''_{m+q,\, m+q+1}.  \label{a5b}
\end{eqnarray}
Thus, the energy spectra of $H'$ and $H''$ coincide, and the whole spectrum
of $H^{full}$ contains two copies of the eigenspectrum of $H^{red} = H'$ (or $H^{red} = H''$).
(The matrix of $H''$ is transformed to $H'$ by changing sign of all its even [or odd] basis functions.)


\end{document}